Electronic Transport in a Three-dimensional Network of 1-D Bismuth Quantum Wires

T. E. Huber, Howard University, Washington, D.C. 20059

and

M. J. Graf, Department of Physics, Boston College, Chestnut Hill, MA 02467

The resistance R of a high density network of 6 nm diameter Bi wires in porous Vycor glass is studied in order to observe its expected semiconductor behavior. R increases from 300 K down to 0.3 K. Below 4 K, where R varies approximately as $\ln(1/T)$, the order-of-magnitude of the resistance rise, as well as the behavior of the magnetoresistance are consistent with localization and electron-electron interaction theories of a one-dimensional disordered conductor in the presence of strong spin-orbit scattering. We show that this behaviour and the surface-enhanced carrier density may mask the proposed semimetal-to-semiconductor transition for quantum Bi wires.

# I. INTRODUCTION

The latest experimental advances in preparation of nanowires have given give rise to both new applications and fundamental science. One of the most exciting prospects is that of an ideal quantum wire of diameter $d$ less than the Fermi wavelength $\lambda_F$. Profound changes in the density of state (DOS) and deep modifications of the transport properties are expected, as seen, for example in single wall carbon nanotubes. Enormous effort has been applied to the study of carbon nanotubes since its discovery in 1991 with considerable success.[1] Comparatively little attention has been given to the semimetal Bi, that has the smallest effective mass among all known materials. Due to this property, quantum confinement in Bi can be observed in nanowires of larger diameter than those of any other material. Here we present a study of the electronic transport in a network of Bi wires of 6 nm- diameter, much smaller than the bulk electronic Fermi wavelength $\lambda_F$ of 25 nm, and we are able to confirm some theoretical predictions.

The quantum-confinement driven semimetal-to semiconductor (SMSC) transition has been studied recently in carbon nanotubes[2] Bulk Bi is also a semimetal whose three-conduction band minima at the L-points overlap the valence-band maximum at the T-point by about 40 meV and a SMSC transition has also been predicted for Bi. Sandomirskii[3] pointed out that as a result of the electrons'zero-point energy, the band overlap decreases for thin films and if the film is thin enough a gap is formed and the semimetal turns into a semiconductor. The critical thickness $t_c$ for this transition for Bi is expected to be 30 nm. Despite many transport and optical investigations of quantum phenomena in thin films the experimental evidence is not conclusive. These films' resistivity is weakly T-dependent, unlike a typical intrinsic semiconductor that exhibits a thermally activated behavior. This has been interpreted in terms of a very short effective mean free path $l \sim t$, and a surface-enhanced carrier density.[4,5] Since the majority carrier concentration then becomes insensitive to the introduction of an energy gap one should expect no abrupt changes at the transition point. The alternative view, that the SMSC does not occur because the boundary condition of vanishing electron wavefunction is spoiled by disorder, also has proponents. Experiments, however, are lacking. Gurvitch[7] and Brandt *et al*[8] have shown that 200 nm diameter Bi wires display metallic behavior down to 2 K and exhibit quantum size effects. More recently, Liu, Chien and Searson presented electrical transport measurements on thick wires ($d > 200$ nm$>>d_c$, $\lambda_F$).[9] There is renewed interest in the electronic properties of quantum wires, of diameter $d$ where $d < \lambda_F$, and where $\lambda_F$ is the Fermi wavelength equal to 25 nm for the electrons (7 nm for holes) in bulk Bi, because theory indicates that they may have outstanding thermoelectric properties.[10,11] Recently, Zhang, Sun, Dresselhaus, Ying and Heremans[12] reported a reversal of the magnetoresistance temperature dependence for Bi wires with 65 nm diameter relative to those with 110 nm diameter. The phenomenon is due to the quantization of the transverse momentum of the carriers which results in the SMSC transition. A nanowire of diameter $d<<d_c$, the critical diameter, is a simpler case. However, to our knowledge the electronic transport of a wire of $d<d_c,\lambda_F$, has not been addressed experimentally. One approach is that of confining Bi in a microporous solid insulator. In this study we take such an approach and we use porous Vycor glass (PVG) because it is a unique prototype monolithic nanoscale porous material. The use of PVG as a confining host has been particularly fruitful, for example, in studying the properties of superfluid helium,[13] in a restricted geometry. The characterization of PVG, particularly the surface, is intrinsically difficult but the basic network structure is reasonable well known from adsorption isotherm and small-angle X-ray scattering studies.[14] Many aspects of the network structure have been clarified by studying the temperature dependence of the the critical field of In injected in PVG. [15]

Confinement introduces scattering and it is known that electronic transport in disordered conductors can be ballistic, diffusive or in a localized regime. Relevant lengthscales set by are the carriers' elastic $l_e$ and inelastic pathlength $l_\phi$. In disordered systems $l_\phi$ can exceed $l_e$ by several orders of magnitude. The coherent superposition of the scattered electron wave results in back-scattering. This causes a weak localization correction to the classical resistance which is composed of the Drude and electron-phonon contributions. Such corrections appear in many metallic systems of restricted geometry (wires and films) which show $d\rho/dT <0$ at low enough temperatures. The resistance rise can be expressed as $\delta R/R \sim \Lambda/L$, where L is the length of a conductor having the characteristic quantum resistance $h/(2e^2) = 12.9$ k$\Omega$. $\Lambda$ has been derived in the case of localization,[16] where $\Lambda = l_\phi$ for H=0, and also for electron-electron interaction[17] theories. The application of these ideas to Bi[18] two-dimensional and three-dimensional films is well known. Bi, being a heavier metal, has a large spin-orbit interaction, characterized by a lengthscale $l_{so}$, which can change the character of the low temperature correction to one of "antilocalization" ($d\rho/dT >0$). We show for the first time that 6 nm-diameter Bi wires exhibit only weak localization and only at the lowest temperatures (T<2 K).

Progress in the study of true one-dimensional quantum wire systems has been slow due to the difficulty of fabricating such materials. On the other hand, the quasi-1D case has been around for 20 years and is well within the realm of microelectronic technology. Mohanty, Jariwala, and Webb[19] have reviewed many studies of this type. All papers cited consider the quasi-1D case $\lambda_F$ <W<$l_\phi$ where W is the wire width. A more recent case is presented by Khavin et al.[20] In this case $\lambda_F = 10$ nm and W is 50 nm. Also, a common technique of confining the two-dimensional electron gas is to pattern a gate. Applying a negative voltage to the gate with a split gate geometry depletes the electron gas beneath it results of a narrow conducting channel; the conducting width can be tuned down to zero by making the gate voltage more negative. However, the actual induced potential is not known directly, and theoretical studies indicate that it has a saddle shape. In the constriction, electrons are confined in the lateral direction and also retarded by the presence of a potential barrier in the wire direction.[21] As we will show, our experiment presents the (real) 1D case, a network of wires of diameter $d<\lambda_F <l_\phi$. In our experiment the confinement is well defined by the wire geometry and the network structure. The physics underlying the two experimental cases should not be confused.

The plan of the paper is as follows. In Sec. II we discuss briefly the sample preparation method, sample characterization and other experimental issues. In Sec. III we analyze the experimental results.

II. SAMPLE PREPARATION AND CHARACTERIZATION

The samples are prepared by melting pure Bi (99.9999 %) and injecting it into porous Vycor glass (PVG) by applying hydrostatic pressures of 5 kilobars. We have used this technique previously for In and other materials.[15] The PVG used has an average pore diameter $d$ of 6 nm. The interconnected network of pores occupies approximately 30% of the total volume. The shiny black samples have approximately 80 % of the pore volume filled with Bi. While the silica backbone structure is complex and interconnected, one can consider it being made up of silica particles of a characteristic size of 26 nm, and this structure is unchanged by the injection process.[22] X-ray diffraction (XRD) from the Bi-PVG composite shows that the Bi retains its rhombohedral (trigonal) structure but with shrinkage of its unit cell. For example, the (102) planes in the hexagonal indexing system, that are separated by 0.328 nm in bulk Bi are separated by 0.320 nm in the composite. This corresponds to a lattice linear contraction of approximately 2.5%. This is likely a result of the injection process since Bi expands on solidification by 3.3 % An estimate for the average Bi crystallite size D can be obtained from the widths of the XRD peaks through Scherrer's equation.[23] A value $D=$ 9 nm was determined from the .017 FWHM of the peak corresponding to the (102) planes, comparable to the PVG average pore diameter $d$=6 nm. A scanning electron micrograph image of the composite is shown in Fig. 1. The Bi in the composite was exposed by etching with HF. We have observed that these samples are not very stable and that samples larger than a few millimeters crumble when subject to stress. Still, we were able to select samples that were robust and therefore suitable for our experiments.

The sample resistance was measured using four terminal d.c. and a.c. (f = 100 Hz) techniques. Electrical contact was made via brass wires attached with silver epoxy to gold pads deposited in a vacuum evaporator. At 300 K the resistivity of the composite is roughly 21 mΩ-cm, and cooling to 4 K increases this to 27 mΩ-cm. The ratio of the room temperature resistivity of the composite to that of the bulk is $r = \rho_{Bi-PVG}/\rho_{bulk} = 175$. A simplified structural model of the composite is that of a 3D simple-cubic lattice of 6-nm diameter nanowires where the periodicity of the network lattice is $s \sim 26$ nm, roughly the size of the silica particles. A value of $r \sim 12$ is estimated by considering that Bi occupies 80% of the pore volume and that about 1/3 of the metal wires are oriented along the current flow. For example, $r$ is 20 for In-filled Vycor at room temperature, where the electronic mean free path $l_e << d$ and surface scattering is not very important.[15] Therefore, $r$ for Bi-PVG is roughly one order of magnitude larger than expected from structural considerations, indicating that the resistivity is much larger than the bulk at room temperature (RT). Accordingly, the RT resistivity, $\rho_{Bi}$, of Bi in the nanowires is estimated to be 1.8 mΩ-cm. We will show later that this value is consistent with results obtained for thin Bi films

It is obvious that Bi is subject to large stresses in the pores of PVG and that these might induce a phase change. A hydrostatic pressure of 20 kBar induces the transition from semimetal Bi(I) to the metallic Bi(II). We can calculate the stresses present in our samples in the following manner: from a linear contraction of 2.5 % in the [102] direction and taking Young's modulus to be $5 \times 10^{11}$ dyn/cm$^2$,[24] we find an stress of $1.3 \times 10^9$ dy/cm$^2$ or 13 kBar. This pressure is insufficient to promote the Bi(I)-(II) transition.[25] Also, the effect of compression in thick (500 nm) Bi films has been investigated and is known that a pressure of 10 kBar increases the film resistivity by roughly 7%. Therefore, we believe the stresses in our samples are not directly relevant to the high resistivity of Bi in the composite.

III. RESULTS

The temperature-dependent resistivity of Bi-PVG is shown in Fig. 2. The inset displays the resistivity over a wide temperature range exhibiting a negative temperature coefficient. In comparison, the bulk Bi resistivity obeys a $T^2$ law at low temperatures and is roughly proportional to T for T > 100 K.[26] The expected semiconductor activation energy can be obtained as follows. The presumed thin film's gap dependence was found theoretically[2] and experimentally[4] to obey roughly $\Delta = 40$ meV $((t_c/t)^2-1)$. Assuming the behavior of a nanowire of $d=$ 6 nm to be that of a thin film of the same thickness we obtain $\Delta = 1$ eV. Clearly the Bi wire network fails to display the thermally activated behavior expected for an intrinsic semiconductor with such a large band gap. This can indeed be understood in terms of a surface-enhanced carrier density. In thin films,[3,4] it was found that $p=p_i+p_s/t$ where $p_i$ is the effective carrier intrinsic concentration resulting from band overlap ($p_i = 2.5 \times 10^{17}$/cm$^3$ for thick films) and $p_s = 3 \times 10^{12}$/cm$^2$ is the effective density per unit film area. At low temperatures ($k_B T < \Delta$) the majority carrier concentration becomes temperature independent. For our nanowires, considering the number of surface states is proportional to the surface area, the effective bulk concentration can be estimated to be $5 \times 10^{18}$/cm$^3$. This concentration is consistent with our results. From $\rho_{Bi}=(nq\mu)^{-1}$ we obtain $\mu=10^3$ cm$^3$/(sec volt), in good agreement with thin Bi film results for $t = 20$ nm. More study is needed to understand this issue and, in particular, the impact of the surface states on the expected thermoelectric properties.[7,8] In order to exhibit the intrinsic regime, the Bi samples should have lower net doping levels, which may be obtained by doping or surface modification. Due to the close analogy between the phenomena exhibited by Bi films and nanowires with regard to the temperature dependence of the resistivity we expect that the two subjects will develop in close association.

Superimposed on the slowly rising $\rho(T)$ as T decreases, a much sharper temperature rise takes place at temperatures below 4K. This is the main focus of this paper and is shown in more detail in Fig. 3. The variation is seen to be approximately logarithmic although the data is also roughly consistent with a $T^{-1/2}$ dependence. This particular temperature dependence is indicative of one-dimensional weak localization in wires. It is useful to compare the order-of-magnitude of our results with those of other workers. Williams and Giordano[27] studied Au wires of 8 nm diameter and found that the resistance fractional rise $[(R(1.5 K)/R(4 K))-1]$ is $1.5 \times 10^{-2}$ and had a $T^{1/2}$ temperature dependence. If this number is multiplied by the factor 64/36 to take into account the smaller diameter of our wires, the extrapolated temperature rise is $2.6 \times 10^{-2}$, in numerical agreement with our result which is $1.7 \times 10^{-2}$. This suggests that our composite sample closely resembles a network of nearly one-dimensional wires. It is interesting that if a simple-cubic three-dimensional Bravais lattice of wires is assumed, the resistance of each 30 nm long section of 6 nm diameter wire is 9 k$\Omega$ at 4 K. This estimate is consistent with the characteristic quantum resistance $h/(2e^2) \sim 12.9$ K$\Omega$ for which weak localization effects are expected to be important.[16] Alternatively, we can look at the low temperature anomaly as a manifestation of disorder in a 3D sample. The Yoffe-Regel-Mott criterion[28] for localization in 3D indicates that the Drude theory is inappropriate for samples with $\rho > 2$ m$\Omega$ $\lambda_F$(in nm). Since the Fermi wavelength $\lambda_F$ can be estimated to be 25 nm for electrons and 7 nm for holes, our sample resistivity of 21 m$\Omega$.cm is clearly problematic.

Williams *et al*[27] and also White *et al*[29] observed that their metallic wires exhibit negligible magnetoresistance (MR). In contrast, our samples show a sizable one (Fig. 4). It should be noted that the MR is positive and that it increases for decreasing temperatures. The transverse[30] MR of the Bi-PVG is quite different from that of the bulk Bi which exhibits a $B^2$ dependence at low magnetic fields and a $B^{1.6}$ dependence[31] in the high magnetic field range. The behavior shown by Bi-PVG is reminiscent to that of thin Bi films where the Coulomb anomaly and the weak localization anomaly occur simultaneously.[4,18]

The MR of quasi-1D wires is well studied. Altshuler and Aronov[32] calculated the localization contribution to the magnetoresistance of wires by treating the magnetic field as a perturbation. This approximation actually improves for decreasing wire diameter. Taking spin orbit effects into account[33] $\Delta R(T,H) \propto 3(1/l_\phi^2 + 1/l_{so}^2 + 1/l_H^2)^{-1/2} - (1/l_\phi^2 + 1/l_H^2)^{-1/2}$, where $l_H$ is the magnetic field dephasing length which is roughly $1.73 \phi_0/H$ and $\phi_0 = 2\pi\hbar c/2e$. The best fit of the MR data at 2.3 K is shown in Fig. 4., using $l_\phi$ = 80 nm and $l_{so}$ = 30 nm. In comparison, $l_\phi$ is 100 nm for thin Bi films at the same temperature.[18] These parameters satisfy the condition $l_{so}, l_\phi \gg d$, where $d$ is the wire diameter. However, since our wires are short and $l_{so}, l_\phi \gg s$ = 26 nm, the magnetoresistance is also modulated by the mesoscopic network structure. For example, quantum interference in the electronic path around a silica particle should lead to a magnetoresistance oscillation with a period $\Delta H = 4\phi_0/(\pi s^2) \sim 3T$. The effect of the network in shaping the magnetoresistance, the relative orientation between wire and magnetic field, as well as the evaluation of the various phase-breaking mechanisms[34] is outside the scope of this paper. It should be noted that weak localization effects were not considered in the interpretation of their Bi nanowire magnetoresistance measurements by the previous workers.[7-11] Because these studies were for larger diameter wires at higher temperatures we can infer that weak localization effects appear in Bi at low temperatures and only for very small diameters wires.

We now discuss some structural considerations. We have already noted that large Bi-PVG samples are not stable and crumble under severe stress. The cracks fill with bulk Bi. We have also observed that samples that appear to be whole will undergo small changes from one experimental run to another. Upon aging of several months the composite resistance becomes slightly smaller, i.e, the resistance at 4 K changes from 26 mΩ.cm to 25.2 mΩ cm. This decrease is accompanied by a shoulder at 2 K in the temperature dependence of the resistivity. The I(V) is non-linear near the bump as indicated by different R(T) readings under d.c. and a.c. excitation. In comparison, the low temperature (T ~ 4 K) I=I(V) curve of "new" samples is linear up to a current density of 0.6 A/cm$^2$. This contrast is consistent with the expectation that the sample is becoming more "granular", and consisting of thinner wires, upon aging. This is expected to be a result of aging since such changes would minimize the surface energy of the composite. We have observed that, upon aging, the magnetoresistance becomes larger. For example, $\Delta R/R$ is 0.05 at 3 T and 0.35 K, while it is 0.02 of the "new" samples. This is consistent with weak localization since thinning down of wires would tend to increase the magnetoresistance.[33] It is observed that the room temperature resistance changes very little in this process, actually decreasing for older samples. In a first approximation, the shortened mean free path ($l_e \sim d$) in smaller wires is compensated by an increased number of majority carriers (n ~ d$^{-1}$).

Weitzel and Micklitz[35] have studied granular systems built of well-defined rhombohedral Bi nanoclusters (d~4 nm) in a variety of matrices (Ge, O$_2$, H$_2$, and Xe). They found that many of these composites exhibit superconductivity at low temperatures (T$_c$~4 K). This behavior contrasts with the localization behavior of our sample. However, the onset of superconductivity may be related to the bump observed by us at 2 K in "aged" or granular samples.

IV. SUMMARY

We present results of an experimental study of the temperature and magnetic-field dependence of the resistivity of a network of 6 nm diameter Bi wires where d<<$\lambda_F$. The wire diameter, 6 nm, is much smaller than the critical diameter for the SMSC transition estimated to be 30 nm. However, we do not observe the expected temperature-activated behavior typical of an intrinsic semiconductor in the temperature range investigated. This can be interpreted in terms of surface-enhanced carriers. Also, we do not observe strong localization and the composite is a basically a good conductor. The observed temperature rise at low temperature and associated magnetoresistance are interpreted in terms of weak localization effects due to disorder. The 80 nm dephasing length scale as inferred from the data is somewhat larger than the network periodicity and the wirelength *s* and, therefore, the magnetoresistance could be modulated by the network mesoscopic structure. Undoubtedly, further study of long, straight nanowires will help understand better the role of localization in electronic transport in networks of 1-D wires; we are currently undertaking such studies, utilizing parallel nanochannels insulating host materials.[36]

The work of T.E.H. was supported by the Army Research Office through DAA H04-95-1-0117 and by the National Science Foundation through DMR-9632819. M.J.G. is supported in part through Research Corporation Grant RA0246.

REFERENCES

1. S. Tans *et al*, Nature <u>386</u>, 474 (1997).

2. G. Lopinski, V.I. Merkulov, and J.S. Lannin, Phys. Rev. Lett. <u>80</u>, 4641 (1998).

3. V.B. Sandormirskii, JEPT <u>25</u>, 101 (1967).

4. Y.F. Komnik *et al*, Sov. J. Low Temp. Phys. <u>1</u>, 51 (1975).

5. C.A. Hoffman *et al*, Phys. Rev. <u>B51</u> 5535 (1995).

6. H.T. Chu, Phys. Rev. <u>B51</u> 5532 (1995).

7. M. Gurtvich, J. Low Temp. Phys. <u>38</u>, 777 (1980).

8. N.B. Brandt, D.V. Gitsu, A.A. Nikolaeva, and Ya.G. Ponomarev, Soviet JETP, <u>24</u> 273 (1976).

9. K.Liu et al, Phys. Rev. <u>B58</u>, R14681 (1998).

10. E.N. Bogacheck, A.G. Scherbakov, and U. Landman, in "Nanowires", edited by P.A. Serena and N. Garcia (Kluwer, Dordrecht, 1997).

11. L.D. Hicks and M.S. Dresselhaus, Phys. Rev. <u>B47</u>, 16631 (1993).

12. Z. Zhang, X.Sun, M.S. Dresselhaus, J. Ying, and J. Heremans, Appl. Phys. Lett. <u>73</u> 1589 (1998).

13. See, for example, C. Lie-Zhao, D.F. Brewer, C. Cirit, E.N. Smith, and J.D. Reppy, Phys. Rev. <u>B33</u> 106 (1986)

14. For a review, see for example, A. Ch. Mitropolous *et al* Phys. Rev. <u>B52</u>, 10035 (1995).

15. M.J. Graf *et al*, Phys. Rev. <u>B45</u>, 3133 (1992). Also, F. Dong *et al*, Sol. State Comm. <u>101</u>, 929 (1997).

16. D.J. Thouless, Phys. Rev. Lett. <u>39</u>, 1167 (1977). P.W. Anderson, D.J. Thouless, E. Abrahams, and D.S. Fisher, Phys. Rev. <u>B22</u>, 3519 (1980).

17. B.L. Altshuler, D. Khmel'nitzkii, A.I. Larkin, and P.A. Lee, Phys. Rev. <u>B22</u>, 5142 (1980).

18. Y.F. Komnic *et al*, Sov. J. Low. Temp. Phys. 7, 656 (1982). Y.F. Kommik *et al*, Solid State Comm. 44, 865 (1982).
19. P. Mohanty, E.M.Q. Jariwala, and R.A. Webb, Phys. Rev.Lett. 78 3366 (1997), Table I.
20. Y.B.Khavin et al, Phys. Rev. Lett. 81 , 1066 (1998).
21. C.W.J. Beenakker and H.van Houten in Solid State Physics, edited by H. Ehrenreich and D. Turnbull (Academic Press, New York, 1991)
22. T.E. Huber, P.W. Schmidt, J.S. Lin, and C.A. Huber, to be published.
23. C.A. Huber in "Handbook of Nanophase Materials", edited by A. Goldstein (M. Dekker, New York, 1996).
24. Y.A. Burenov *et al*, Sov. Phys. Sol. State 14, 215 (1972).
25. F.P. Bundy, Phys. Rev. 110, 314 (1958).
26. V.N. Lutskii, JEPT 25, 101 (1967).
27. G.D. Williams and N. Giordano, Phys. Rev. 33 8146 (1986).
28. Y. Imry in "Introduction to Mesoscopic Physics" (Oxford University Press, Oxford, 1997).
29. A.E. White, M. Tinkham, W.J. Skocpol, and D.C. Flanders, Phys. Rev. Lett. 48 1752 (1982).
30. Although the macroscopic current is transverse to the magnetic field, the nanowires are oriented at random. Since the MR has different H-dependence in the parallel and transverse geometries (Ref. 17) our sample geometry will affect the observed H-dependence of the composite MR.
31. L.A. Falkovskii, Soviet Phys. Uspekii 94, 1 (1968).
32. B.L. Altshuler and A.G. Aronov, JEPT Lett. 33, 515 (1981).
33. S. Wind, M.J. Rooks, V. Chandrasekhar, and D.E. Prober, Phys. Rev. Lett. 57, 633 (1986).
34. M.E. Gershenson et al, Phys. Rev. B51 10256 (1995).
35. H.B. Weitzel and H. Micklitz, Phys. Rev. Lett. 66, 385 (1991).

36. T.E. Huber, M.J. Graf, and C.A. Foss, in *Thermoelectric Materials- The Next Generation Materials for Small-Scale Refrigeration and Power Generation Applications* edited by T. M. Tritt, H.B. Lyon, Jr., G. Mahan, and M.G. Kanatzidis (Materials Research Society, Pittsburg, 1999), p.227.

FIGURE CAPTIONS

Fig. 1.  Scanning electron micrograph of the Bi-PVG composite. Dark areas correspond to silica.

Fig. 2.  The resistivity from 0.3 K to room temperature.

Fig. 3.  Low temperature resistivity of Bi-injected PVG.

Fig. 4  Magnetoresistance of Bi-injected PVG at 0.3 K. The solid line is a fit to the theory of Ref. 33.

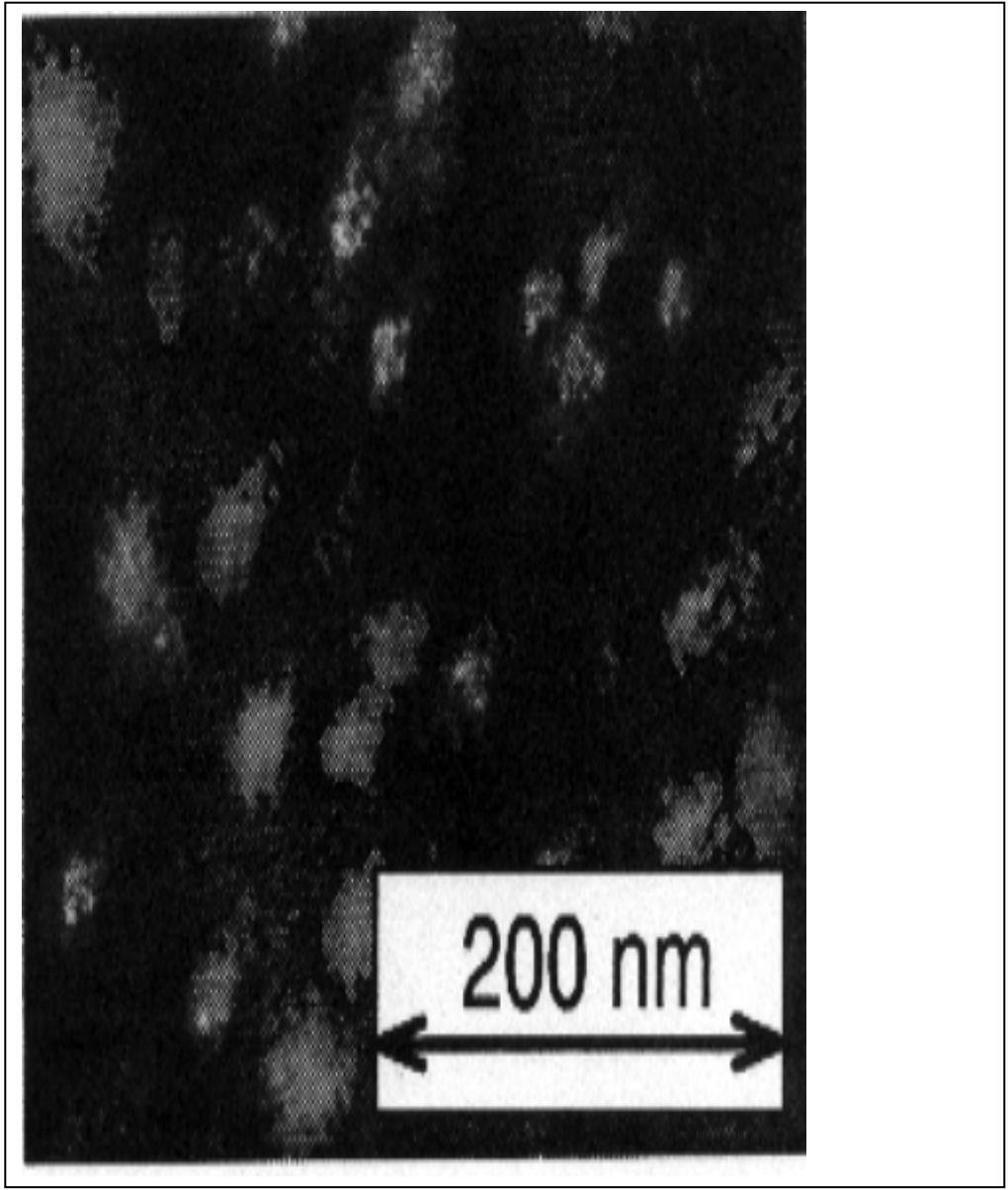

Fig. 1

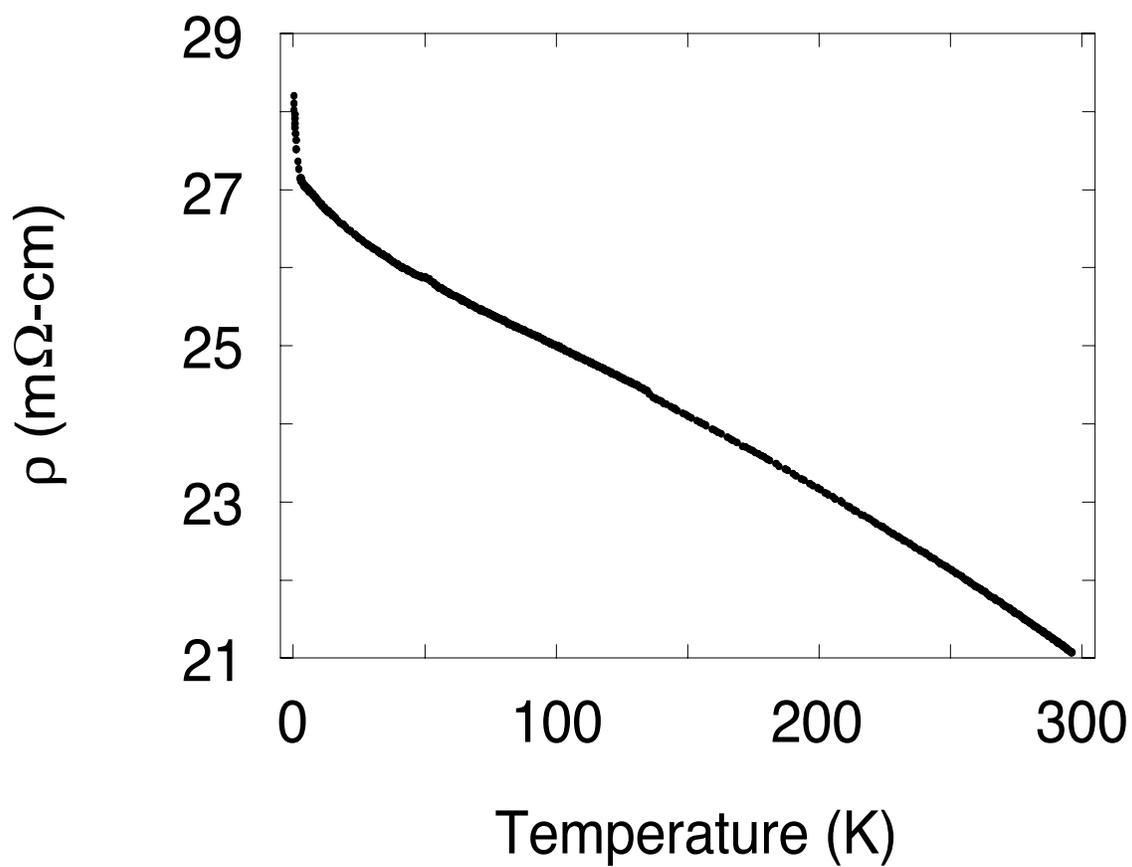

Fig. 2

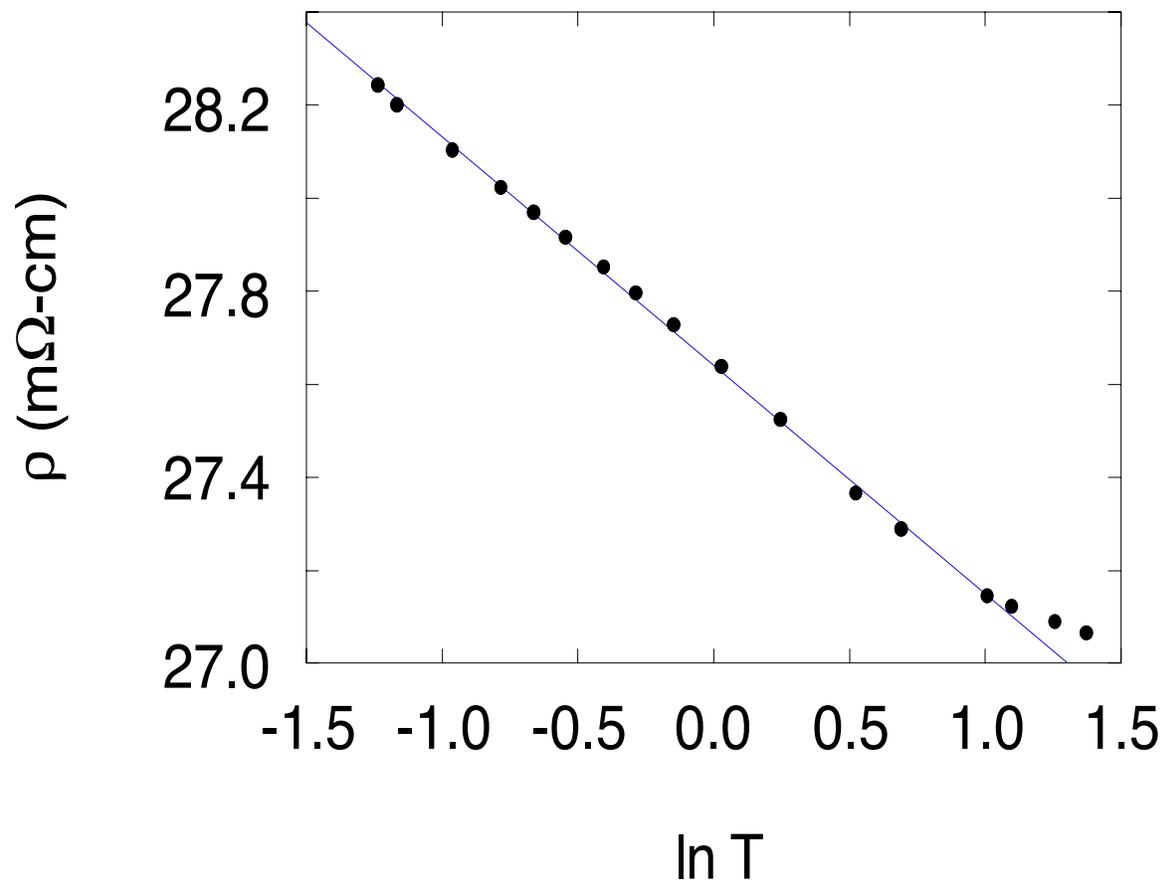

Fig. 3

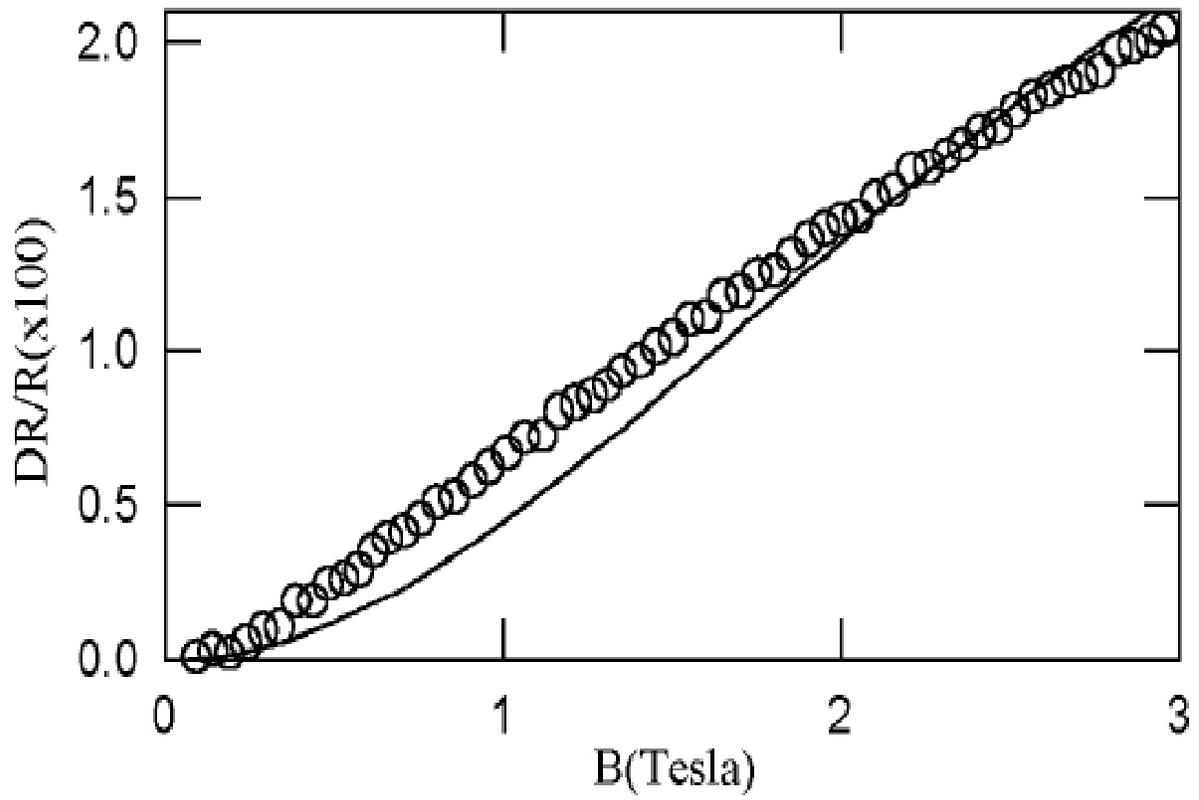

Fig. 4